\documentclass[twocolumn,aps,prl,superscriptaddress,floatfix]{revtex4}
\usepackage[T1]{fontenc}
\usepackage[utf8]{inputenc}
\usepackage[english]{babel}
\usepackage{color}
\usepackage{float}
\usepackage{braket}
\usepackage{amsmath}
\usepackage{amstext}
\usepackage{amssymb}
\usepackage{graphicx}
\usepackage{bbold}
\usepackage[unicode=true,pdfusetitle,
bookmarks=true,bookmarksnumbered=false,bookmarksopen=false,
breaklinks=false,pdfborder={0 0 1},backref=false,colorlinks=false]
{hyperref}
\newcommand{\angstrom}{\textup{\AA}}
\hypersetup{
	colorlinks,linkcolor=blue,citecolor=blue,urlcolor=blue}
\usepackage{soul}
\makeatletter
\date{}

\makeatother
\setul{0.5ex}{0.3ex}
\definecolor{Blue}{rgb}{0,0.0,1}
\setulcolor{Blue}
\usepackage{babel}

\begin{document} 

\author{Tarik P. Cysne}
\affiliation{Instituto de F\'\i sica, Universidade Federal Fluminense, 24210-346 Niter\'oi RJ, Brazil} 
\email{tarik.cysne@gmail.com}

\author{Marcio Costa}
\affiliation{Instituto de F\'\i sica, Universidade Federal Fluminense, 24210-346 Niter\'oi RJ, Brazil} 

\author{Luis M. Canonico}
\affiliation{Catalan Institute of Nanoscience and Nanotechnology (ICN2), CSIC and BIST, Campus UAB, Bellaterra, 08193 Barcelona, Spain}

\author{M. Buongiorno Nardelli}
\affiliation{ Department of Physics and Department of Chemistry, University of North Texas, Denton TX, USA}

\author{R. B. Muniz}
\affiliation{Instituto de F\'\i sica, Universidade Federal Fluminense, 24210-346 Niter\'oi RJ, Brazil}

\author{Tatiana G. Rappoport}
\affiliation{Instituto de Telecomunicações, Instituto Superior Tecnico, University of Lisbon, Avenida Rovisco Pais 1, Lisboa, 1049001 Portugal}	
\affiliation{Instituto de F\'\i sica, Universidade Federal do Rio de Janeiro, C.P. 68528, 21941-972 Rio de Janeiro RJ, Brazil}

\title{Disentangling orbital and valley Hall effects in bilayers of transition metal dichalcogenides}

\begin{abstract}
It has been recently shown that monolayers of transition metal dichalcogenides (TMDs) in the 2H structural phase exhibit relatively large orbital Hall conductivity plateaus within their energy band gaps, where their spin Hall conductivities vanish \cite{Us2, OHE_Bhowal_1}. However, since the valley Hall effect (VHE) in these systems also generates a transverse flow of orbital angular momentum it becomes experimentally challenging to distinguish between the two effects in these materials. The VHE requires inversion symmetry breaking to occur, which takes place in the TMD monolayers, but not in the bilayers.  We show that a bilayer of 2H-MoS$_2$ is an orbital Hall insulator that exhibits a sizeable OHE in the absence of both spin and valley Hall effects. This phase can be characterised by an orbital Chern number that assumes the value $\mathcal{C}_{L}=2$ for the 2H-MoS$_2$ bilayer and $\mathcal{C}_{L}=1$ for the monolayer, confirming the topological nature of these orbital-Hall insulator systems. Our results are based on density functional theory (DFT) and low-energy effective model calculations and strongly suggest that bilayers of TMDs are highly suitable platforms for direct observation of the orbital Hall insulating phase in two-dimensional materials. Implications of our findings for attempts to observe the VHE in TMD bilayers are also discussed.  \end{abstract}
\maketitle

{\it Introduction:} The orbital Hall effect (OHE) is the orbital analog of the spin Hall effect and consists in the appearance of a transverse current of orbital angular momentum that is induced by a longitudinally applied electric field~\cite{OHEBernevig}. Recently, a  renewed interest in orbital magnetism and other orbital effects~\cite{OrbitalMagnetism_TBG, Orbital-Magnetism-Go, Murakami-Orb-edelstein, Exp_Orbital_Textures} gave origin to various theoretical studies on the OHE and related phenomena \cite{OHEmetals, Orbital-Rashba_PRB_2013, OrbitalTexture, Fei-Xue_2020, OrbitalTextureBorophene, Us1, Us2, phonon_OH_Park,OHE_Bhowal_1, Propose_OH_Detection}.  The possibility of using the OHE to generate orbital torque in magnetic materials~\cite{Go-spin-orbital-torque, Go-Orbital-torque} motivated new experimental works on orbital dynamics in magnetic multilayers \cite{Exp_orbital_Torque, Exp_orbitalTransport_heterostructures}, raising expectations that orbital angular degrees of freedom may be eventually employed to process information in logic and memory devices. 

The interrelation between the OHE and the presence of orbital textures in reciprocal space ~\cite{OrbitalTexture} has been established and characterised both theoretically and experimentally in several low-dimensional materials~\cite{Exp_Orbital_Textures, Orbital-Texture_Exp_Nature2020, OrbitalTextureBorophene, Us1,Us2, OHE_VHE_PRB_Imaging}, widening the class of systems that may be utilised for orbitronic applications. More specifically, the occurrence of relatively large OHE has been predicted in the 2H structural phase of TMD monolayers \cite{Us2, OHE_Bhowal_1}, where it is associated with the presence of a Dresselhaus-like orbital texture around the valleys \cite{Us2}. However, it is experimentally challenging to observe just the OHE in 2H-TMD monolayers due to the concurrent presence of the VHE that also contribute to the transport of orbital angular momentum in these systems \cite{VHE-Momentum_2}.

It is noteworthy though that the VHE manifests only in the absence of inversion symmetry, which naturally happens for the monolayers, but for bilayers, comprising two monolayers rotated by $\pi$ with respect to each other, the inversion symmetry is restored.  This substantially affects valley related phenomena \cite{Valley-pol-experiment, TMD-bilayer-Exp_1, Exp_GateTuning_VHE}. For instance, the valley Hall conductivity has opposite signs in each layer, cancelling the VHE for the bilayer \cite{Low-Energy-bilayer, Bukard-VHE-bilayer, Fabian_KP_TMDs}, as we shall subsequently discuss. Nevertheless, it is also possible to break inversion symmetry in the bilayers by applying an electric field perpendicular to the layers, by means of which one can control the valley polarisation \cite{ValleyPol-TMD-bilayer} and the VHE intensity \cite{Exp_GateTuning_VHE} with a gate voltage.

Here we perform calculations of the orbital Hall conductivities for ultra-thin films (single layer and bilayer) of 2H-MoS$_2$ which is representative of this class of systems. We combine Density Functional Theory (DFT) and an effective low-energy model to disentangle the valley and orbital physics of TMD bilayers and explore some of their topologic orbital features.

Implications of our findings regarding interpretations of recent experiments on the electric control of the VHE in MoS$_2$ bilayers ~\cite{Exp_GateTuning_VHE, TMD-bilayer-Exp_1} are 
also briefly discussed.  Our results strongly indicate that bilayers of TMDs constitute a fertile play-ground for exploring orbital angular momentum current generation in 2D-like systems.    

{\it DFT results:}
Our DFT calculations~\cite{DFT1,DFT2} were performed with the plane-wave-based code \textsc{Quantum Espresso}~\cite{QE-2017}. The exchange and correlation potential is treated within the generalised gradient approximation (GGA)~\cite{PBE}. The ionic cores were described with fully relativistic projected augmented wave (PAW) potentials~\cite{PAW}. We used a cutoff energy of 63 Ry for the wavefunctions and a value 10 times larger for the charge density. In order to reproduce the interlayer distance of the MoS$_{2}$ bilayer we have used the DFT-D3~\cite{DFT-D3} method, which describe reasonably well the van der Walls forces in these systems. We have chosen a 10$\times$10$\times$1 reciprocal space sampling, and to avoid spurious interaction due to periodic boundary conditions we insert a vacuum spacing of 15\AA. We constructed an effective tight-binding Hamiltonian from our DFT calculations using the pseudo atomic orbital projection (PAO) method~\cite{PAO1,PAO3}. The PAO method consists of projecting the DFT Kohn-Sham orbitals into the compact subspace spanned by the pseudo atomic orbitals which are naturally built-in into the PAW potentials. The PAW potentials used for the Mo and S were constructed with a $sspd$ and $sp$ basis, respectively. 

Once the PAO Hamiltonian is constructed we can calculate the spin Hall (SH) and orbital Hall (OH) responses to an applied electric field \cite{OHEorigins, GraphaneOHE, OHEmetals, Mele_PRL_2019, OHEBernevig, Us1, Us2}. Up to linear order on the external field they are given by: 
\begin{eqnarray}
\sigma^{\eta}_{OH(SH)}=\frac{e}{(2\pi)^2}\sum_{n} \int_{BZ} d^2k f_{n\vec{k}}~\Omega_{n,\vec{k}}^{X_{\eta}},
\label{conductivity}
\end{eqnarray}
where $\sigma^{\eta}_{OH(SH)}$ is the orbital Hall (spin Hall) DC conductivity with polarisation along the $\eta$-direction, and
\begin{eqnarray}
\Omega_{n,\vec{k}}^{X_{\eta}}= 2\hbar\sum_{m\neq n}\text{Im} \Bigg[ \frac{\big<\psi_{n,\vec{k}}\big|j_{y,\vec{k}}^{X_{\eta}}\big|\psi_{m,\vec{k}}\big>\big<\psi_{m,\vec{k}}\big|v_x(\vec{k})\big|\psi_{n,\vec{k}}\big>}{(E_{n,\vec{k}}-E_{m,\vec{k}}+i0^+)^2}\Bigg] \label{Kubo2}
\end{eqnarray}
represents the angular-momentum-weighted Berry curvature~\cite{OrbitalTexture, Mele_PRL_2019}. Here, $E_{n,\vec{k}}$ denotes the eigenvalue of the Hamiltonian $H(\vec{k})$ in reciprocal space, and $|\psi_{n,\vec{k}}\big>$ is the corresponding eigenvector; $n$ is the band index, $\vec{k}$ is the wave vector. The velocity operators are defined as $v_{x(y)}(\vec{k})=\partial H(\vec{k})/\partial \hbar k_{x(y)}$, where x and y specify the Cartesian axes, and we assume that the electric field is applied along the $\hat{x}$ direction. The current density operator component along $\hat{y}$ with polarisation $\eta$ is defined as $j_{y,\vec{k}}^{X_{\eta}}=\big(X_{\eta}v_y(\vec{k})+v_y(\vec{k})X_{\eta} \big)/2$, where for the SH conductivity $X_{\eta}=\hat{s}_\eta$ and for the OH conductivity (OHC) $X_{\eta}=\hat{\ell}_\eta$; $\hat{s}_\eta$ and $\ell_\eta$ represent the $\eta$-components of the spin and of the atomic angular momentum operators, respectively. This is implemented in the \textsc{Paoflow} code~\cite{PAO5} that has been successfully used to study topological materials~\cite{Costa2019,hoti}, and time dependent spin dynamics~\cite{fegete} among other topics.  For our conductivity calculations we have increased the sampling to 200$\times$200$\times$1 k-points in the 2D B.Z.

\begin{figure}[h]
	\centering
	 \includegraphics[width=0.9\linewidth,clip]{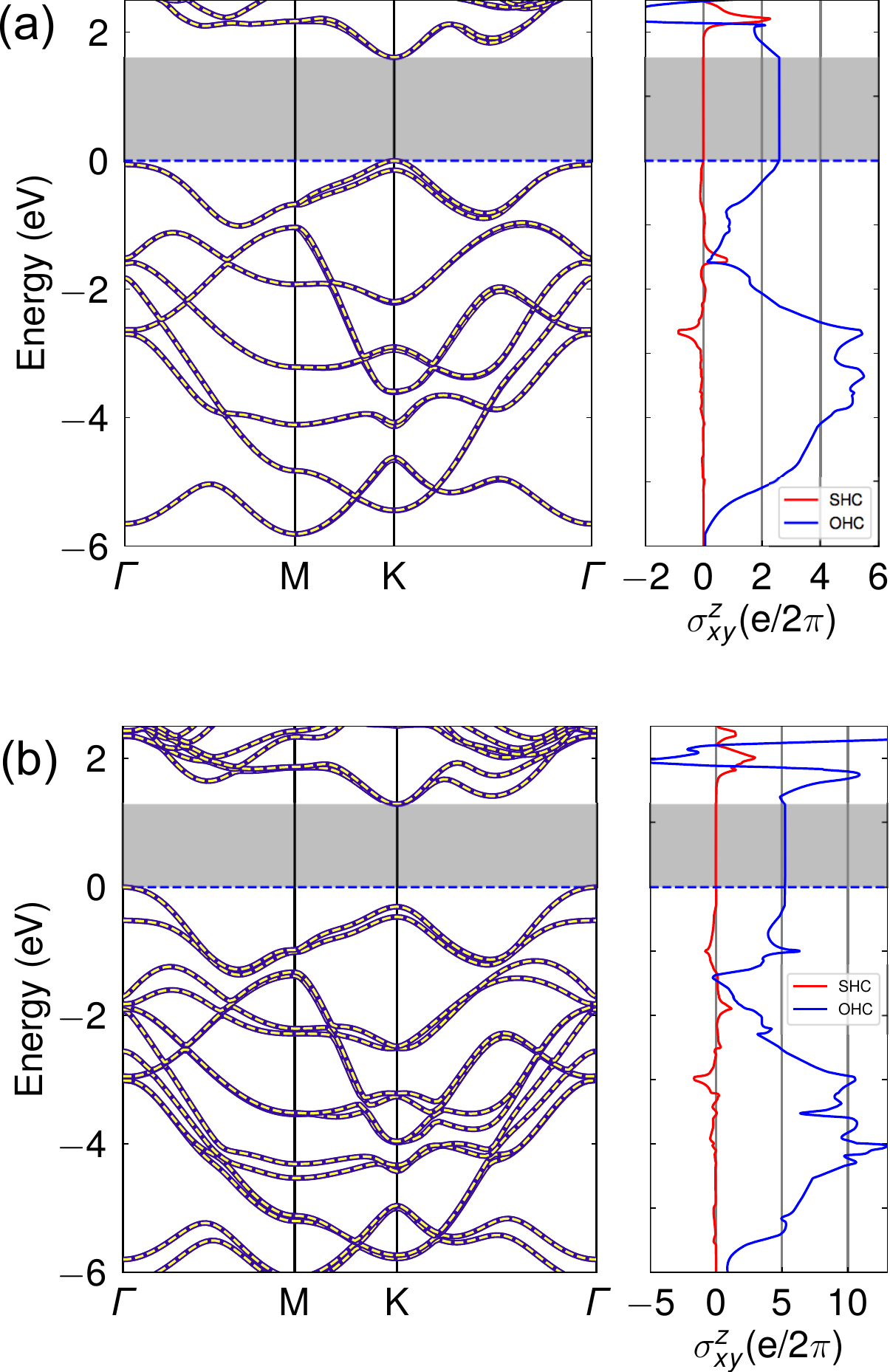}   
  
	\caption{Energy band structures (left panel) together with the spin-Hall and orbital-Hall conductivities (right panel) calculated for a MoS$_{2}$ monolayer (a) and for a MoS$_{2}$ bilayer (b). The purple solid and yellow dashed lines depict the DFT and \textsc{Paoflow} band structure calculations, respectively. The horizontal blue dashed line shows the Fermi level.}
	\label{fig:fig1}
\end{figure}

Figures~\ref{fig:fig1} (a) and (b) illustrate results of our calculations for a monolayer and for a bilayer of 2H-MoS$_2$, respectively. In the left panels of Figure~\ref{fig:fig1} we compare the band structures obtained from DFT (purple solid lines) and from \textsc{Paoflow} (yellow dashed lines). The agreement between the two approaches is excellent. For the monolayer we obtain a direct energy-band gap of 1.60 eV, whereas for the bilayer we found an indirect gap of 1.28 eV, in agreement with previous calculations~\cite{MoS2-DFT}. The results for the SHC (red solid lines) and for the OHC (blue lines) are shown in right-hand side panels of Figure~\ref{fig:fig1}. In accordance with our previous results for the monolayer \cite{Us2} we note in right panel of Figure~\ref{fig:fig1} (a) that the SHC vanishes in the main energy gap, but the OHC is finite and exhibits relatively high plateau of $\approx 2.6$ ($e/2\pi$) in height within this energy range. For the bilayer, however, the right-hand side panel of panel Fig.~\ref{fig:fig1}(b) show that the height of OHC plateau is essentially twice the monolayer value, while the SHC remain null in the main energy gap because it is topologically trivial. Although other regions in the B. Z. contribute to the OHC~\cite{OHE_Bhowal_1}, the main contribution originates from the orbitally projected Berry curvature in the vicinity of K and K', as illustrated in the supplementary material~\cite{SM}. To explore the physics behind these results, it is instructive to make use of a low-energy approximation around the $K$-points (valleys) of the BZ,  to build a simple model that is able to reasonably describe the main transport characteristics of these systems. 

{\it Low energy calculations:}
Similarly to the monolayers, the low-energy physics of TMD bilayers is dominated by the $d_{z^2}$, $d_{x^2-y^2}$ and $d_{xy}$ atomic orbitals of the transition metal atoms \cite{Xiao_PRL_2012, ThreebandXiao, Low-Energy-bilayer}. We follow references \onlinecite{Bukard-VHE-bilayer} \onlinecite{Low-Energy-bilayer} to build a simplified tight-binding (TB) model Hamiltonian in reciprocal space, which is expanded up to first order in the electronic momentum around the valleys located at $\vec{K}=(4\pi/3a)\hat{x}$ and $\vec{K}'=-\vec{K}$. 
This procedure leads to the following Hamiltonian:

\begin{eqnarray}
\tilde{H}(\vec{q}_\tau)=\begin{bmatrix}
\Delta & \gamma_+  & 0 & 0\\
  \gamma_- & -\tau s_z \lambda & 0 & t_{\perp}\\
0 & 0 & \Delta & \gamma_-\\
0 & t_{\perp} & \gamma_+ & \tau s_z \lambda
\end{bmatrix},
\label{Heff}
\end{eqnarray}
where  $\gamma_\pm = at(\tau q_x \pm i q_y)$, $\tau=\pm 1$ is the valley quantum number associated with valleys $K$ and $K'$, respectively. Here, $\vec{k} = \vec{q} + \tau \vec{K}$ where $\vec{q}$ represents the wavevector relative to valleys and $s_z$ denotes the usual Pauli matrix. For a 2H-MoS$_2$ bilayer, an archetypal TMD, $\Delta=1.766 \text{eV}$ is the monolayer band-gap, $a=3.160 \angstrom$ is the lattice constant, $t=1.137 \text{eV}$ is the intra-layer nearest-neighbor hopping, $\lambda=0.073 \text{eV}$ is the spin-orbit coupling, and $t_{\perp}=0.043 \text{eV}$ is the interlayer hopping~\cite{Low-Energy-bilayer}. 

The TB basis for this minimal model comprises $\{ \big|d^1_{z^2}\big>,\big( \big|d^1_{x^2-y^2}\big>-i\tau \big|d^1_{xy}\big> \big)/\sqrt{2}, \big|d^2_{z^2}\big>, \big( \big|d^2_{x^2-y^2}\big>+i\tau \big|d^2_{xy}\big> \big)/\sqrt{2}\}$, where the superscripts 1 and 2 specify the two layers of the bilayer, respectively. It is noteworthy that the orbital angular momentum (OAM) operator in this representation is given by $L_z=\text{diag}(0,-2\hbar \tau,0,2 \hbar\tau)$, which clearly does not commute with the Hamiltonian defined in Eq.(\ref{Heff}).

\begin{figure}[h]
	\centering
	 \includegraphics[width=0.99\linewidth,clip]{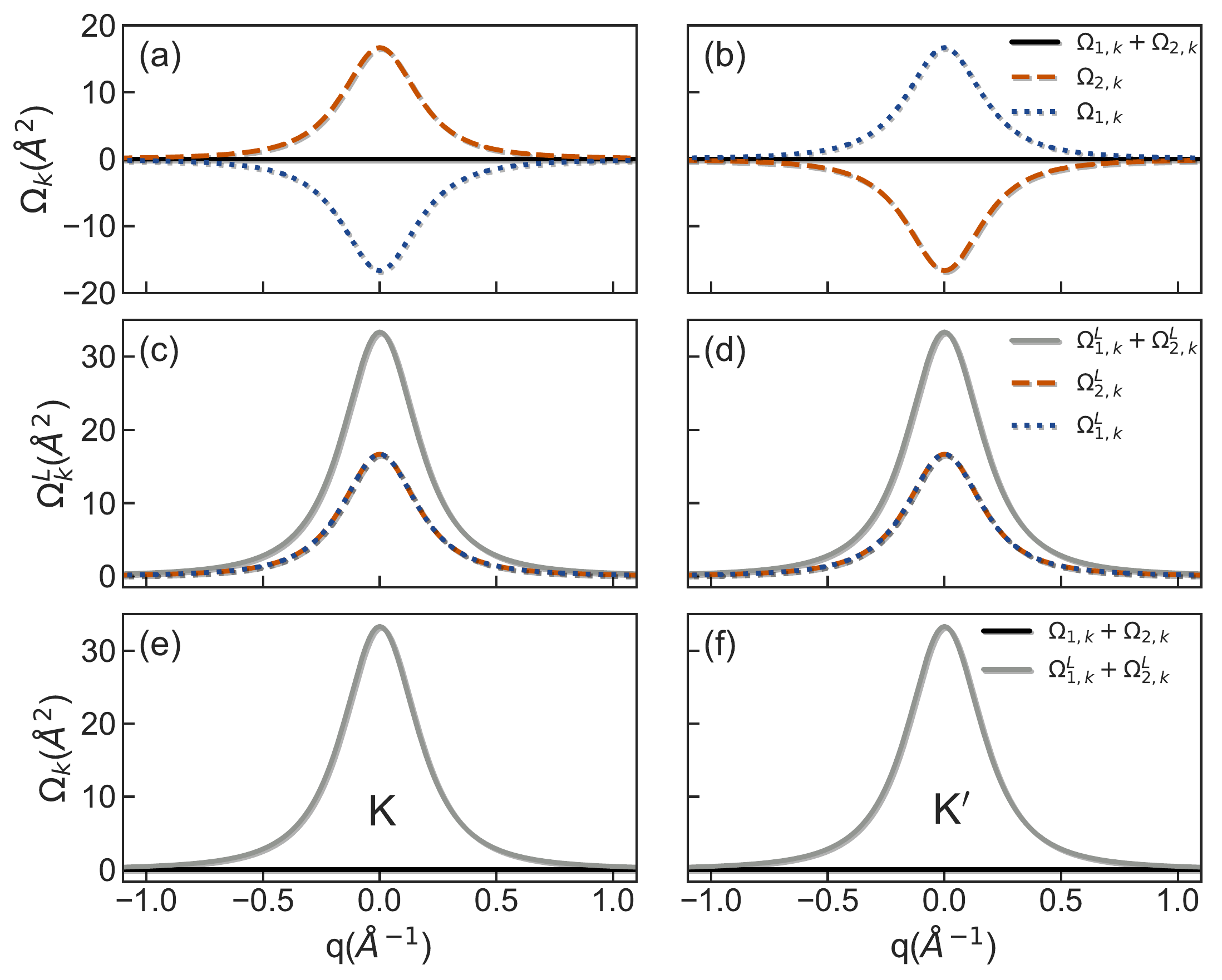}   
  
	\caption{Berry curvature $\Omega_{n,k}$ at points $K$ (a) and $K'$ (b), orbital weighted Berry curvature $\Omega^L_{n,k}$ at points $K$ (c) and $K'$ (d) for the two valence bands $E_{1}(q)$ and $E_2(q)$, associated to the two layers. (e) Total Berry curvature $\Omega_{1,k}+\Omega_{2,k}$ and  (f) orbital weighted Berry curvature $\Omega^L_{1,k}+\Omega^L_{2,k}$ for the bilayer TMD.}
	\label{fig:fig2}
\end{figure} 

Eq. (\ref{Kubo2}) can be used with the four-band low-energy Hamiltonian given by Eq. (\ref{Heff}) to define the Berry and the orbital-weighted Berry curvatures that encode information of the VHE and OHE, respectively. For simplicity, we shall initially neglect the effect of spin-orbit coupling ($\lambda$), thereby restricting Eq. (\ref{Heff}) to a spinless Hamiltonian, and including a degenerescence factor $g_s=2$. Eq. (\ref{Kubo2}) for the orbital weighted Berry curvature may also be employed to calculate the usual Berry curvature $\Omega_{n,k}$, provided that $X_{\eta}$ is replaced by $\hbar\mathbb{1}$. The spinless Hamiltonian generates two valence bands ($E_1(q)$ and $E_2(q)$) that can be regarded as arising from each of the TMD layers because of the relatively small interlayer hopping. Figures \ref{fig:fig2} (a) and (b) present the Berry curvatures for both $E_1$ and $E_2$ calculated around the $K$ and $K'$ points, respectively. The Berry curvature for $E_2$ has a positive peak at $K$ and a negative peak at $K'$, which gives rise to a VHE. The opposite occurs for the Berry curvature of $E_1$, which has a negative peak around $K$ and a positive peak at  $K'$, giving origin to a VHE with an inverted sign. By adding the contributions of both layers, the net Berry curvature is zero in both valleys, and the VHE vanishes. This is a consequence of time-reversal symmetry and the presence of spatial inversion symmetry in the bilayer \cite{Bukard-VHE-bilayer, Xiao_PRL_2012, Bilayer-TMD-symmetries}. A similar situation occurs for TMDs with the $T$ and $T'$ structural phases, such as WTe$_2$ \cite{Tony-low-centrosimmetric-TMD}. Figures \ref{fig:fig2} (c) and (d) show the orbital weighted Berry curvatures for both bands around the $K$ and $K'$ points, respectively. In contrast with the previous case, the peaks of the orbital-weighted Berry curvatures for both bands have the same sign around both valleys. Hence, the total orbital-weighted Berry curvature has a finite value, which leads to an OH insulating phase \cite{Us2} with no VHE, as Figures \ref{fig:fig2} (e) and (f) illustrate. We note that in order to assess just the OHE it is crucial from the experimental point of view to have OHE without VHE, because the VHE also leads a transverse angular momentum current \cite{VHE-Momentum_1,VHE-Momentum_2, VHE-Momentum_3} that is hard to be distinguished from the one generated by the OHE, as it happens for TMD monolayers \cite{Us2}. Thus, our results show that bilayers of 2H-TMDs are very promising candidates for observing the orbital Hall insulating phase with no interferences from VHE or SHE.

\begin{figure}[h]
	\centering
	 \includegraphics[width=0.99\linewidth,clip]{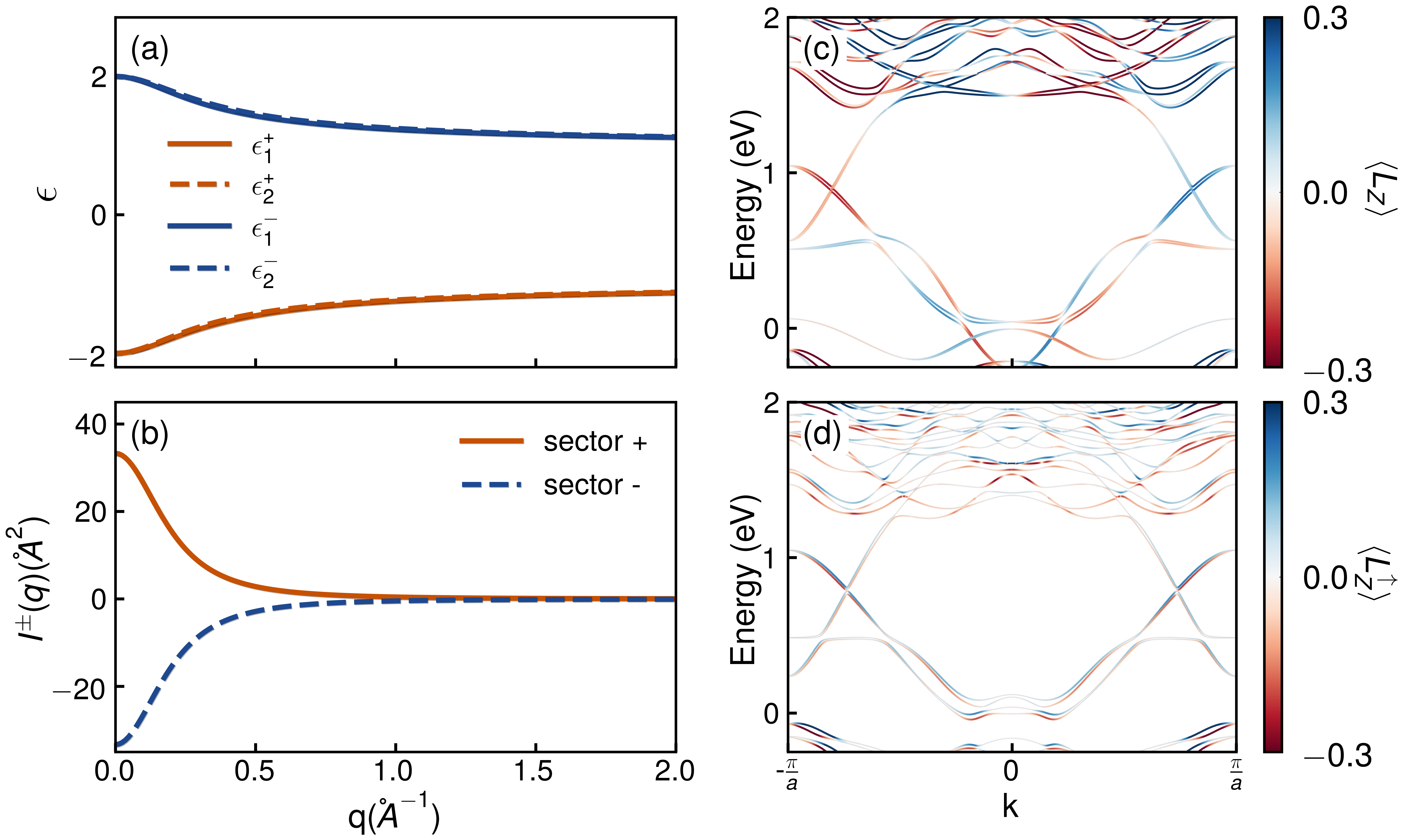}   
  
	\caption{(a) Positive (orange) and negative (blue) eigenvalues $\epsilon$ of the matrix $\mathbb{L}^{\text{v}}(\vec{k})$ calculated as functions of the wavevector amplitude $q$ relative to valleys.  (b)  Integrands of Eq. (\ref{orbChern}) $I^\pm (q) = \sum_{n,\tau} F^{\pm}_{n,\tau}(q)$ calculated as functions of $q$. Band structures of zigzag nanoribbons with 14.8Å in breadth calculated with the \textsc{Paoflow} Hamiltonian for a monolayer (c) and for the spin-up sector of a bilayer (d) of 2H-MoS$_2$. The color code indicates the orbital angular momentum expectation value $\langle L_{z} \rangle$. For better visualization, in the bilayer case only the spin-up bands are showed. The spin-down band-structure are showed in the supplementary material~\cite{SM}.}
	\label{fig:fig3}
\end{figure}
  
Let us now address the topological characterisation of the OH insulating phase in TMD-bilayers. Our Berry curvature analysis suggests that it is possible to associate an orbital Chern number to describe the distinctive nature of these states in analogy with the well known spin Chern number \cite{Kane-Mele_2005, Ezawa_sCN}. Here the situation is slightly more subtle because the operator $L_z$ does not commute with the Hamiltonian of Eq. (\ref{Heff}) for finite $\vec{q}$. This is similar to the problem of a quantum spin Hall insulator in the presence of a Rashba SOC. To address this issue, we follow the procedure developed in Refs. \onlinecite{Prodan-ChernNumber-2009, Sheng-ChernNumber_2011, Sheng-Isospin-Chern-number} to define the orbital Chern number $\mathcal{C}_L$ for the insulating phase of Hamiltonian given by Eq. (\ref{Heff}).  In this formalism, $\mathcal{C}_{L}=(\mathcal{C}_L^+-\mathcal{C}_L^-)/2$ where $\mathcal{C}_L^\pm$ are the Chern numbers calculated with the eigenstates of an OAM operator projected on the valence-band states $\left(\mathbb{L}^\text{v}(\vec{k})=P(\vec{k}){L_z}P(\vec{k})\right)$, where $P(\vec{k})$ is the projector operator. If the bands have orbital polarisation, the spectrum of $\mathbb{L}^{\text{v}}(\vec{k})$ consists of two groups of eigenvalues ($\epsilon$) associated with $m_l=\pm2$ that are symmetrically separated by a gap. The projectors on the eigenstates associated with the positive and negative eigenvalues can then be used to calculate the Chern numbers $\mathcal{C}_L^{\pm}$.
 
Thus, to calculate $\mathcal{C}_{L}$, it is necessary to decompose the valence-band states into two sectors with respect to operator $L_z$. For that purpose, we first obtain the matrix $\mathbb{L}^{\text{v}}(\vec{k})$, with matrix elements given by  $\big<\psi_{n,\vec{k}}\big|L_z\big|\psi_{m,\vec{k}}\big>$, where $n,m$ label the valence-band eigenstates of the low-energy Hamiltonian; more details are given in the accompanying Supplementary Material~\cite{SM}. It is worth mentioning that hereafter we reinstate the spin degree of freedom and the spin-orbit interaction in the Hamiltonian (\ref{Heff}). Figure \ref{fig:fig3} (a) show the eigenvalues of $\mathbb{L}^{\text{v}}(\vec{k})$ calculated as functions of $q$. We clearly see that the spectrum splits in two separated sectors, allowing us to use the eigenstates of $\mathbb{L}^{\text{v}}(\vec{k})$ in each valley $\big|\Phi_{n,\tau}^{\pm}(\vec{q})\big>$ to calculate the Chern numbers:

\begin{eqnarray}
\mathcal{C}_L^{\pm}=\frac{1}{2\pi}\int d^2q \sum_{n,\tau} F^{\pm}_{n,\tau}(q),
\label{orbChern}
\end{eqnarray}
where $F^{\pm}_{n,\tau}(q)=-2 \text{Im} [\big<\partial_{q_x}\Phi_{n,\tau}^{\pm}(\vec{q})\big|\partial_{q_y}\Phi_{n,\tau}^{\pm}(\vec{q})\big>]$. Fig. \ref{fig:fig3} (b)  shows the integrands of Eq. (\ref{orbChern}). Since they have azimuthal symmetry, the calculations of $\mathcal{C}_L^{\pm}$ involve numerical integrations of one-dimensional radial functions only. Our results 
for the insulating phases of the 2H-MoS$_2$ bilayer and single layer are $\mathcal{C}_{L}=2$ and $\mathcal{C}_{L}=1$, respectively, supporting the idea that the relatively weak interlayer hopping in the bilayer makes it behave approximately as a mere superposition of its two constituent monolayers, which are rotated by $\pi$ with respect to each other.

The existence of a nontrivial orbital Chern number should lead to the appearance of edge states when the bulk material is cut to form a ribbon. It is well known that zigzag TMD ribbons present crossing edge-states with interesting orbital properties, even though $\mathbb{Z}_2=0$. \cite{IOPEdgeMoS2}. Figures \ref{fig:fig3} (c) and (d) show the energy band spectra of 2H-MoS$_2$ zigzag nanoribbons, calculated with the use of  \textsc{Paoflow} Hamiltonian for a monolayer and a bilayer  including the orbital angular momentum expectation value $\langle  L_z \rangle$(k) for each eigenstate. ~\cite{Go2020Rashba}. 
The energy band spectrum for a monolayer ribbon depicted in Figure \ref{fig:fig3} (c) clearly shows two pairs of orbitally polarized intra-valley edge states~\cite{Zhang2013} - one for each spin sector- which is compatible with the orbital Chern number $\mathcal{C}_{L}=1$. 
 Results for the bilayer ribbon are displayed in Figure \ref{fig:fig3} (d), where we see two pairs of intra-valley edge states per  spin-sector - which is also compatible with $\mathcal{C}_{L}=2$.  For the bilayer, the presence of inversion symmetry is translated in the existence of positive and negative $L_z$ edge states in both valleys. For clarity, Figure \ref{fig:fig3} (d) presents the results for spin-up while the SM presents the two components.   

{\it Experimental signatures:} Let us now briefly discuss the experimental signatures of the OHE in TMD bilayers. Typically, to characterise the OHE in these materials, one needs the same experimental setups conceived to analyse VHE in TMD bilayers, where inversion symmetry breaking is induced by a gate voltage~\cite{Exp_GateTuning_VHE, Wu2019}. For the bilayer in this case, both the OHE and the VHE lead to magnetic moment accumulation at the sample's edges. To provide some insights into what should be expected in such experiments, we include a gate potential in in Eq. (\ref{Heff}) given by $H_U=\text{diag}(U,U,-U,-U)$. 

\begin{figure}[h]
	\centering
	 \includegraphics[width=0.99\linewidth,clip]{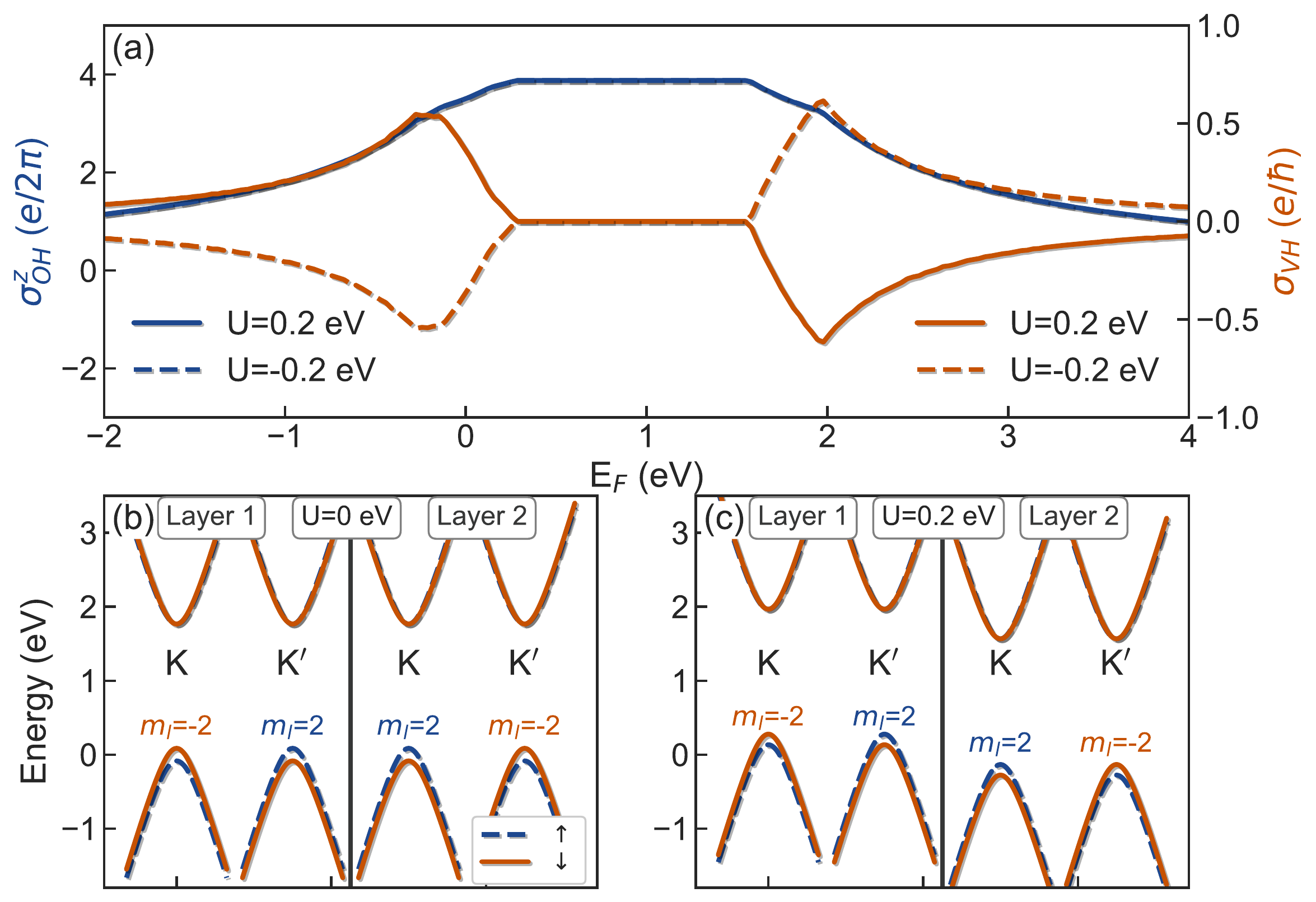}   
  
	\caption{(a) Orbital (blue line) and valley Hall (orange line) conductivities as a function of the Fermi energy $E_F$ for U=0.2 eV (solid line) and U=-0.2 eV (dashed line). Schematic representation of the low energy spectrum for U=0 eV (b)  and $U$=0.2 eV (c).  The orbital polarisation of the top valence bands is indicated by $m_l=\pm2$ and the solid orange and dashed blue lines indicate the two spin orientations.  }
	\label{fig:fig4}
\end{figure}

For a finite $U$, the inversion symmetry is broken in the bilayer and the VHE takes place. The OH and VH conductivities can be calculated using Eqs.(\ref{conductivity}-\ref{Heff}). To calculate the VH conductivity (VHC), we substitute the integrand of Eq.(\ref{conductivity}) by $\Omega_{n,\vec{q}}/\hbar$, and rewrite $\sigma_{VH}=(\sigma_{\tau=+1}-\sigma_{\tau=-1})$.  Fig. \ref{fig:fig4}(a) shows our results for the OHC ($\sigma^z_{OH}$) and VHC ($\sigma_{VH}$) calculated as functions of the Fermi energy ($E_F$) for positive and negative values of $U$. There are clear diferences between the two quantities. While the OHC is an even function of $E_F$, the VHC is odd. Also, the OHE is an even function of $U$, whereas the VHE is odd - the valley magnetic moment inverts when $U$ changes sign~\cite{TMD-bilayer-Exp_1}. Panels (b) and (c) of Fig. \ref{fig:fig4}  show the energy spectrum for $U=0$ and $U=0.2$ eV, respectively. It is clear that $U$ produces a rigid energy-band shift for the two layers, without changing their orbital polarisations. The OHE should remain unchanged for small variations of $E_F$, but decreases when $E_F$ crosses any band, as individual bands in each valley contribute to the total Chern number. 
We have also performed DFT calculations for the OHE in the presence of an electric field applied perpendicularly to the layers~\cite{FMauri}. The results support our low-energy analysis and are presented in the SM~\cite{SM}.

Kerr rotation microscopy experiments\cite{Exp_GateTuning_VHE} showed that a bilayer of MoS$_2$ exhibits a sizeable Kerr rotation even in the absence of an applied gate voltage. It was argued that this unexpected behaviour could originate from substrate induced inversion symmetry breaking. Recent non-local resistance measurements in hBN encapsulated bilayer of MoS$_2$ also exhibited non-local signal at zero gate voltage~\cite{Wu2019}. The interpretation was the same, although one should not expect hBN to cause such a large inversion symmetry breaking effect. On the other hand, the OHE could be the source of this experimental evidence and explain the unexpected signals at zero bias in the bilayers. Careful experimental analysis of Kerr rotation and non-local resistance measurements as functions of gate voltage may help to distinguish between the orbital and valley Hall effects in these materials. The results illustrated in Fig. \ref{fig:fig4}, in light of the experiments reported in Refs. \cite{Exp_GateTuning_VHE, Wu2019}, suggest that ultrathin films of TMDs are promising platforms to explore the OHE in 2D materials.

{\it Final remarks and conclusion:} Our DFT calculations showed that centrosymmetric two-dimensional materials, such as a bilayer of 2H-MoS$_2$, can host an orbital Hall insulating phase in the absence of both spin and valley Hall effects. Using MoS$_2$ as a prototype of the TMD family, we have also unveiled the topological nature of OHE in these systems and calculated the orbital Chern numbers for 2H-TMDs.  Our work clarifies the interplay between orbital and valley Hall conductivity in bilayer TMDs. We found that, in the absence of a gate voltage between the layers, the magnetic moment accumulation observed in experiments should be dominated by the OHE as VHE is zero in centrosymmetric materials. For finite bias, OHE and VHE are still decoupled and can behave as competing effects.

\begin{acknowledgments}
	We acknowledge CNPq/Brazil, CAPES/Brazil, FAPERJ/Brazil and INCT Nanocarbono for financial support.  TGR acknowledges funding from Fundação para a Ciência e a Tecnologia and Instituto de Telecomunicações - grant number UID/50008/2020 in the framework of the project Sym-Break. She thankfully acknowledges the computer resources at MareNostrum and the technical support provided by Barcelona Supercomputing Center (FI-2020-2-0033). MC acknowledge the National Laboratory for Scientific Computing (LNCC/MCTI, Brazil) for providing HPC resources of the SDumont supercomputer. MBN also acknowledges the High Performance Computing Center at the University of North Texas and the Texas Advanced Computing Center at the University of Texas, Austin, for computational resources. 	LMC is supported by Project MECHANIC (PCI2018-093120) funded by the Ministerio de Ciencia, Innovación y
Universidades. ICN2 is funded by the CERCA Programme/Generalitat de Catalunya and supported by the Severo Ochoa Centres of
Excellence program, funded by the Spanish Research Agency (grant number SEV-2017-0706).

\end{acknowledgments}	

\newpage
\begin{widetext}
\begin{center}
{\bf Supplementary material for ``Disentangling orbital and valley Hall effects in bilayers of transition metal dichalcogenides''}	
\end{center}

\section{MoS$_{2}$ zigzag nanoribbons band structure - PAOFLOW}
\begin{figure}[h]
	\centering
	 \includegraphics[width=0.47\linewidth,clip]{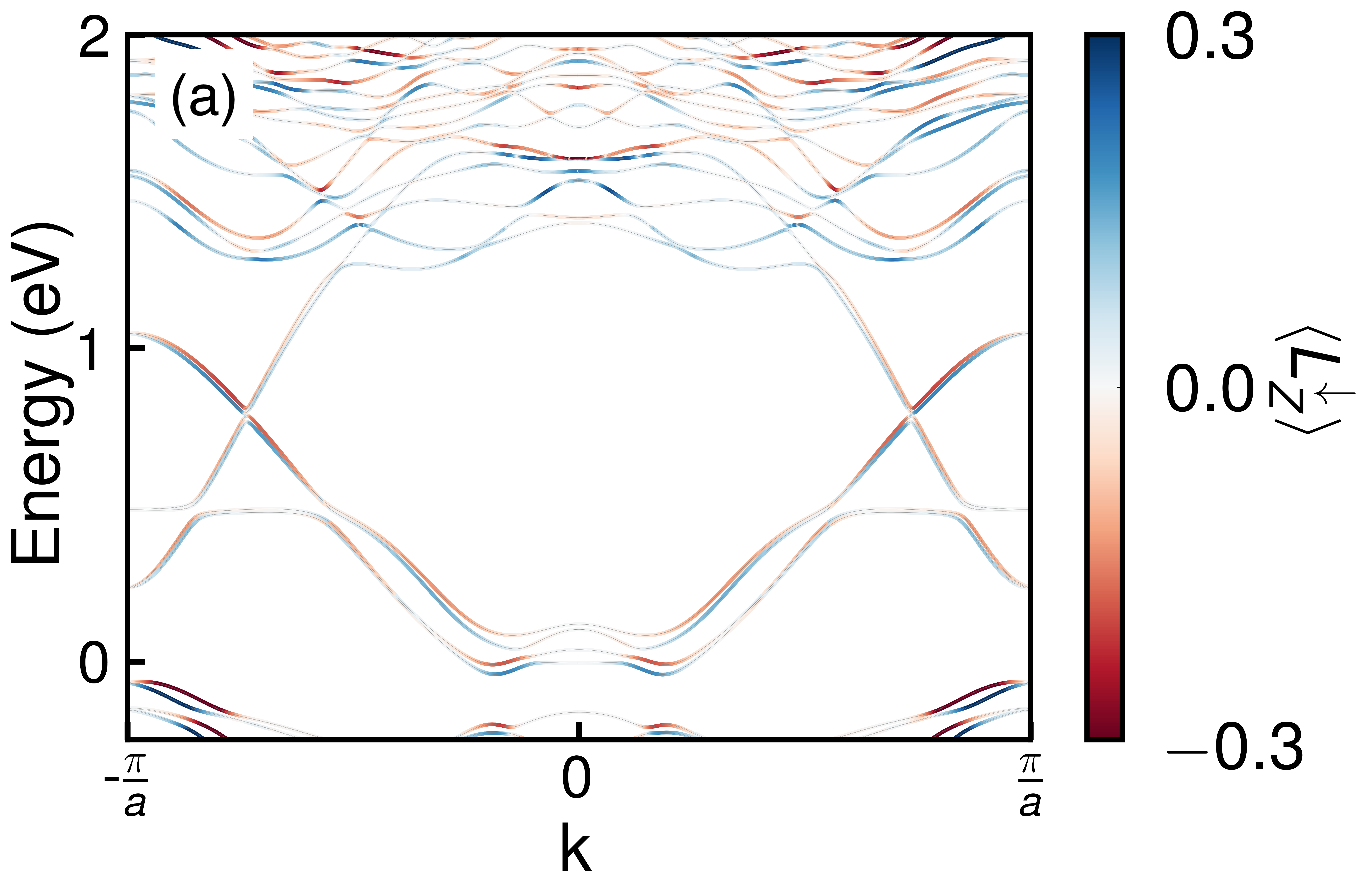}  
	  \includegraphics[width=0.47\linewidth,clip]{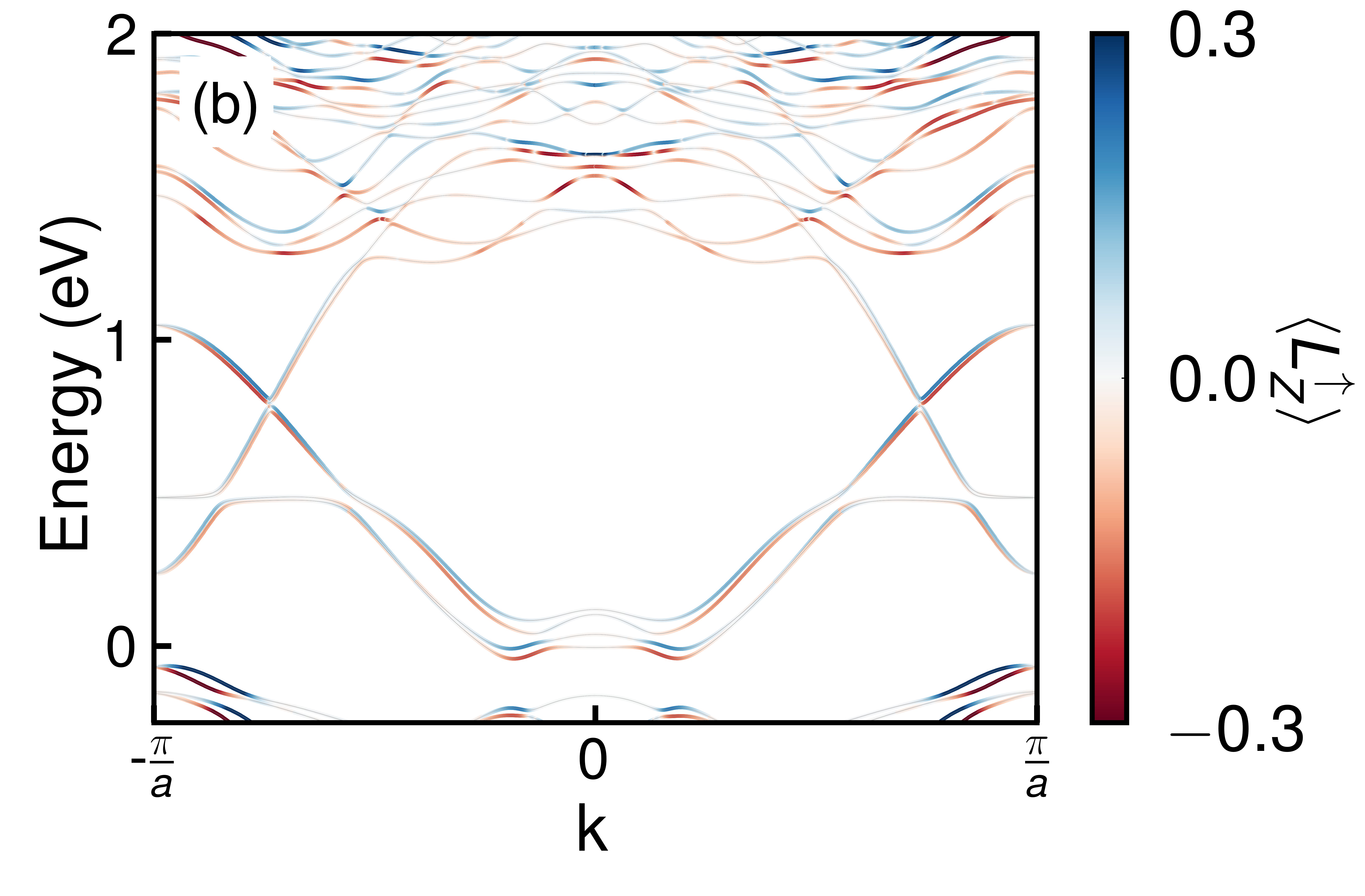}   
  
	\caption{Band structure of a zigzag nanoribbon of 2H-MoS$_2$ bilayer with width of 14.8\AA in breadth calculated with the use of the \textsc{Paoflow} Hamiltonian. Panels (a) and (b) depict the results for the $\downarrow$ and $\uparrow$ spin components, respectively. The color code indicates the orbital angular momentum expectation value $\langle L_{z} \rangle$ for the $\downarrow$ spin bands.}
	\label{fig:nanoribbon}
\end{figure} 

\section{MoS$_{2}$ zigzag nanoribbons band structure - 3 bands approximation}
 \begin{figure}[h!]
	\centering
	 \includegraphics[width=0.5\linewidth,clip]{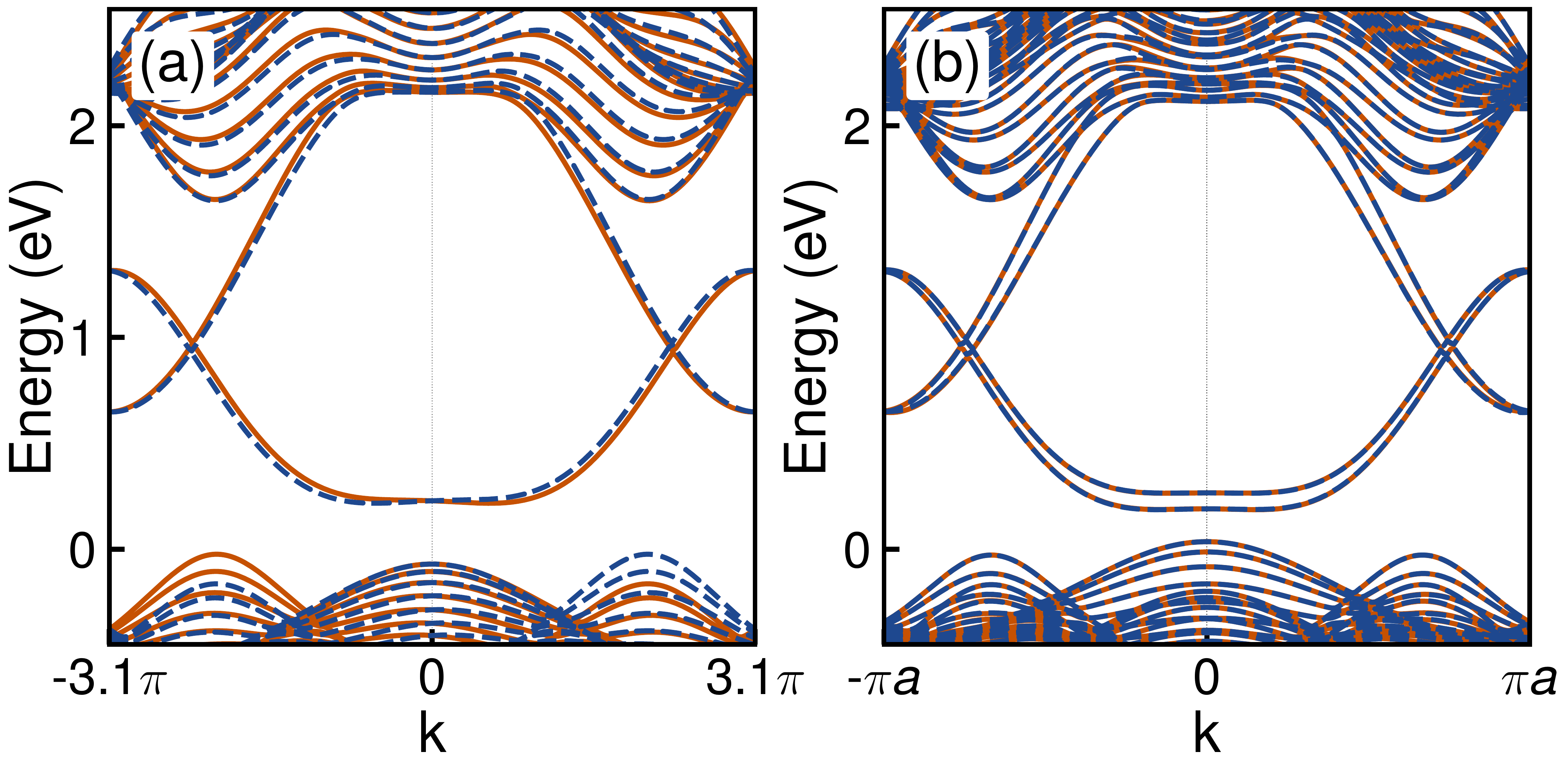}   
	\caption{Zigzag nanoribbon band-structures calculated for a monolayer (a) and for a bilayer (b) of 2H-MoS$_2$ using a simplified 3-bands approximation. Solid orange and dashed blue lines indicate the two spin orientations. }
	\label{fig:nanoribbon3band}
\end{figure} 

Zigzag TMD ribbons exhibit crossing edge-states that can also be modelled by an effective three-bands model, which captures their essential transport features \cite{ThreebandXiao}. Figures \ref{fig:nanoribbon3band} (a) and (b) illustrate the energy band spectra of 2H-MoS$_2$ zigzag nanoribbons, calculated for a monolayer and for a bilayer, respectively~\cite{pybinding,kite}, using the three-bands model of Ref. \onlinecite{ThreebandXiao} and nearest neighbour interlayer hopping integrals only. It is noteworthy that this simplified model does not capture the behaviour of high and low energy bands, because it does not take into account the orbitals $d_{xz}$ and $d_{yz}$ of the transition metal, and treats the effects of the chalcogens (S) perturbatively only. For this reason, the edge-states generated by this model are separated from the bulk valence bands by a non-realistic energy gap. In more realistic descriptions of a TMD nanoribbon  \cite{IOPEdgeMoS2, Emilia_Mos2_2017}, this gap is filled by states with orbital character that are not considered in this simplified model. Nevertheless, the simplified model describes very well the nature of edge-states near their crossing.

\section{\textit{Ab Initio} calculations in the presence of an applied electric field}
\begin{figure}[h]
	\centering
	 \includegraphics[width=0.7\linewidth,clip]{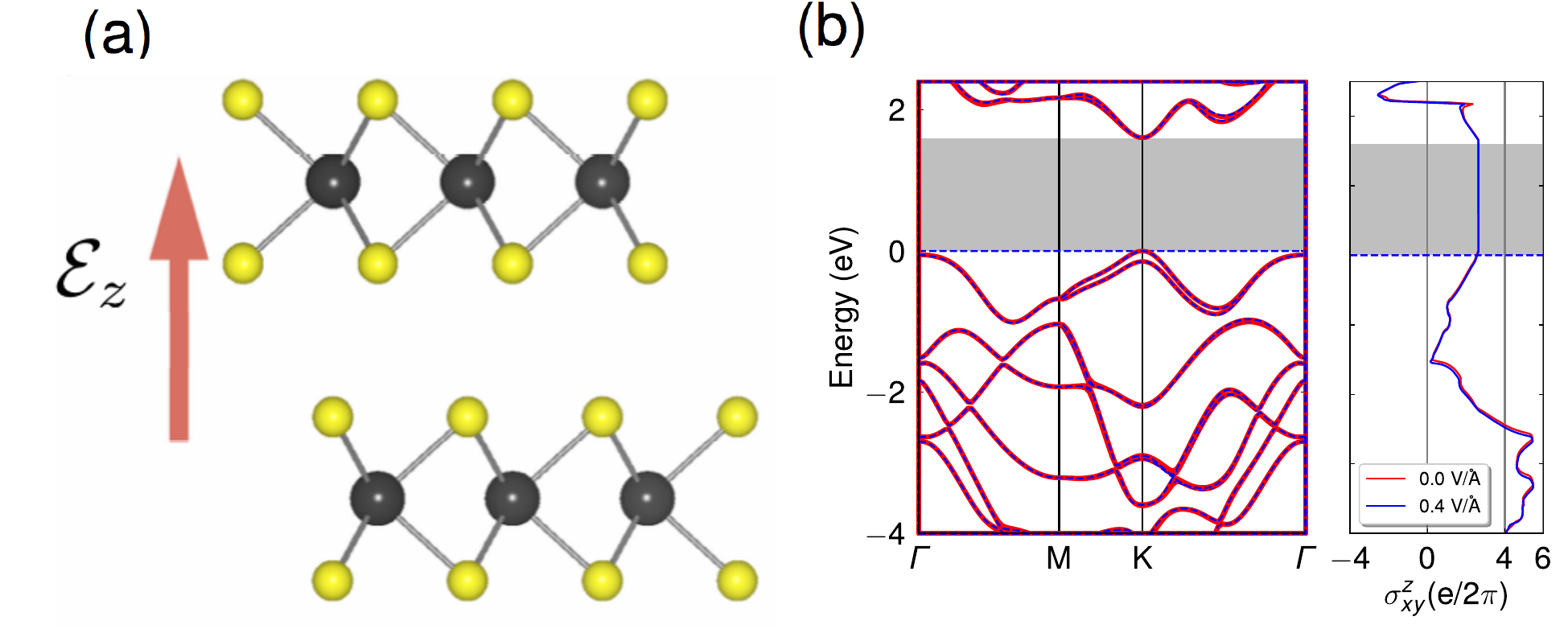}   
  
	\caption{(a) Lateral view of the MoS$_{2}$ bilayer structure. The red arrow reprsents the electric field $\bold {E} = \mathcal{E}_z \hat{z}$. (b) MoS$_{2}$ monolayer energy band structure (left panel) together with the orbital-Hall conductivity (OHC) (right panel) calculated for $\mathcal{E}_{z}=0.0$ V/\AA~(red) and $\mathcal{E}_{z}=0.4$ V/\AA~(blue). The horizontal blue dashed line shows the Fermi level. }
	\label{fig:fig1s}
\end{figure} 

\begin{figure}[h!]
	\centering
	 \includegraphics[width=0.8\linewidth,clip]{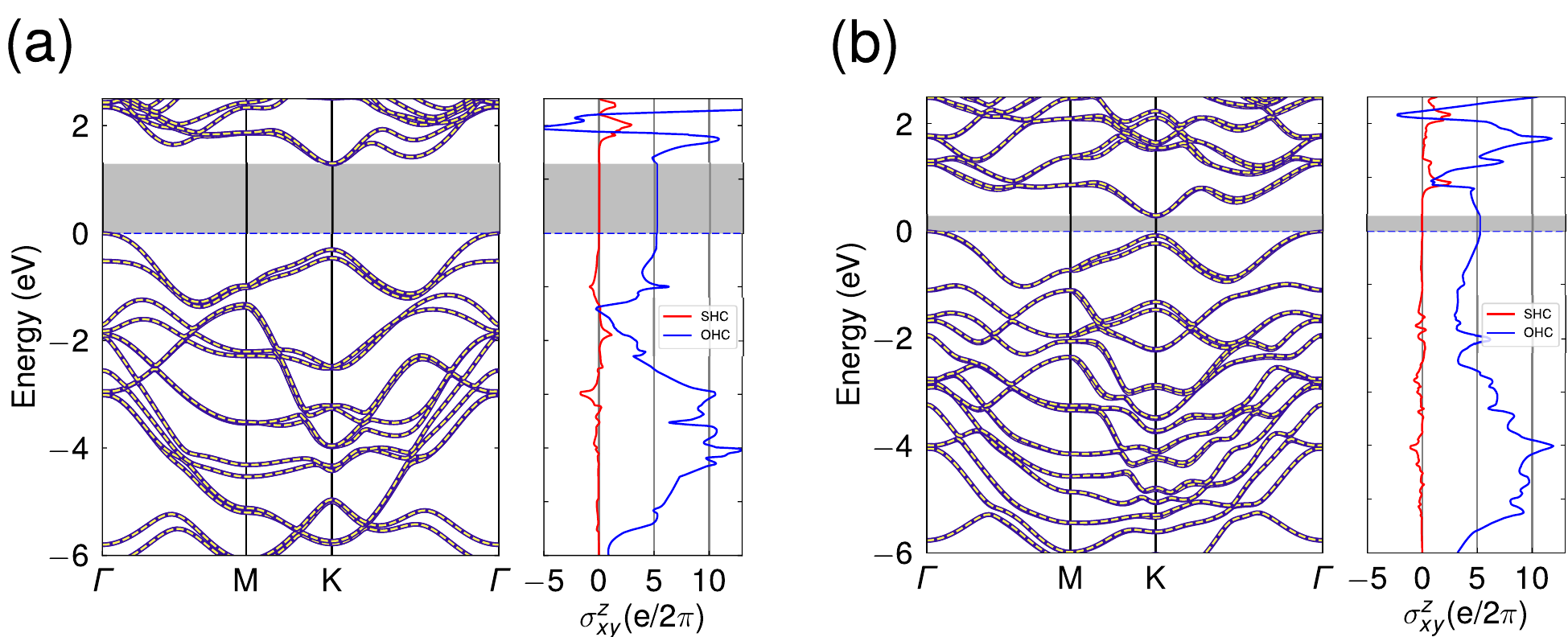}   
  
	\caption{Energy band structures (left panel) together with the spin-Hall and orbital-Hall conductivities (right panel) calculated for a bilayer of MoS$_{2}$ for (a)  $\mathcal{E}_{z}=0.0$ V/\AA~and (b) $\mathcal{E}_{z}=0.2$ V/\AA. The purple-solid and yellow-dashed lines depict the DFT and \textsc{Paoflow} band structure calculations, respectively. The horizontal blue-dashed line denotes the Fermi energy.}
	\label{fig:fig2s}
\end{figure} 

We examine the effects on the electronic structures of the TMD ultrathin films caused by a homogeneous electric field $\bold{E} = \mathcal{E}_z \hat{z}$~\cite{FMauri}, applied perpendicularly to the film layers, using the modern theory of polarization~\cite{Vanderbilt,Umari}. A non local energy functional is defined as the regular energy functional subtracted by the product $\mathcal{E}_z P_z$, where $P_z$ is is the polarization component along the $\hat{z}$-direction. Once the charge density and wave functions are converged we obtain the \textsc{paoflow} Hamiltonian and calculate the electronic properties of interest. In Fig.~\ref{fig:fig1s} we show the band structure together with the OHC of a monolayer of MoS$_{2}$ calculated for $\mathcal{E}_{z}=0$ and $\mathcal{E}_{z}$=0.4 V/\AA. We note that the MoS$_{2}$ monolayer band gap and OHC are virtually not affected by the applied electrical field, which is in agreement with previous results~\cite{MoS2-efield}.

Fig.~\ref{fig:fig2s} shows the energy band structures together with the spin-Hall and orbital-Hall conductivities of a bilayer of MoS$_{2}$ calculated for $\mathcal{E}_{z}=0.0$ V/\AA and $\mathcal{E}_{z}=0.4$ V/\AA. 
We note that the perpendicularly applied electric field breaks the inversion symmetry between the two layers causing a substantial band gap reduction as $\mathcal{E}_{z}=0.4$ increases, whereas the OHC remains practically unchanged. We expect this behavior not to change as long as the band gap remais finite. For applied electric field intensities $0.0 < \mathcal{E}_{z} < 0.4$ V/\AA, the bilayer band energy gap varies almost linearly with $\mathcal{E}_{z}$, as Fig.~\ref{fig:fig3s}(a) illustrates, in agreement with previous DFT calculations~\cite{MoS2-efield}. Figure.~\ref{fig:fig3s}(b) shows the calculated OHC in the same range of electrical field. There is small reduction, $\approx$ 5\%, in the OHC for $\mathcal{E}_{z} < 0.4$ V/\AA.

\begin{figure}[h]
	\centering
	 \includegraphics[width=0.5\linewidth,clip]{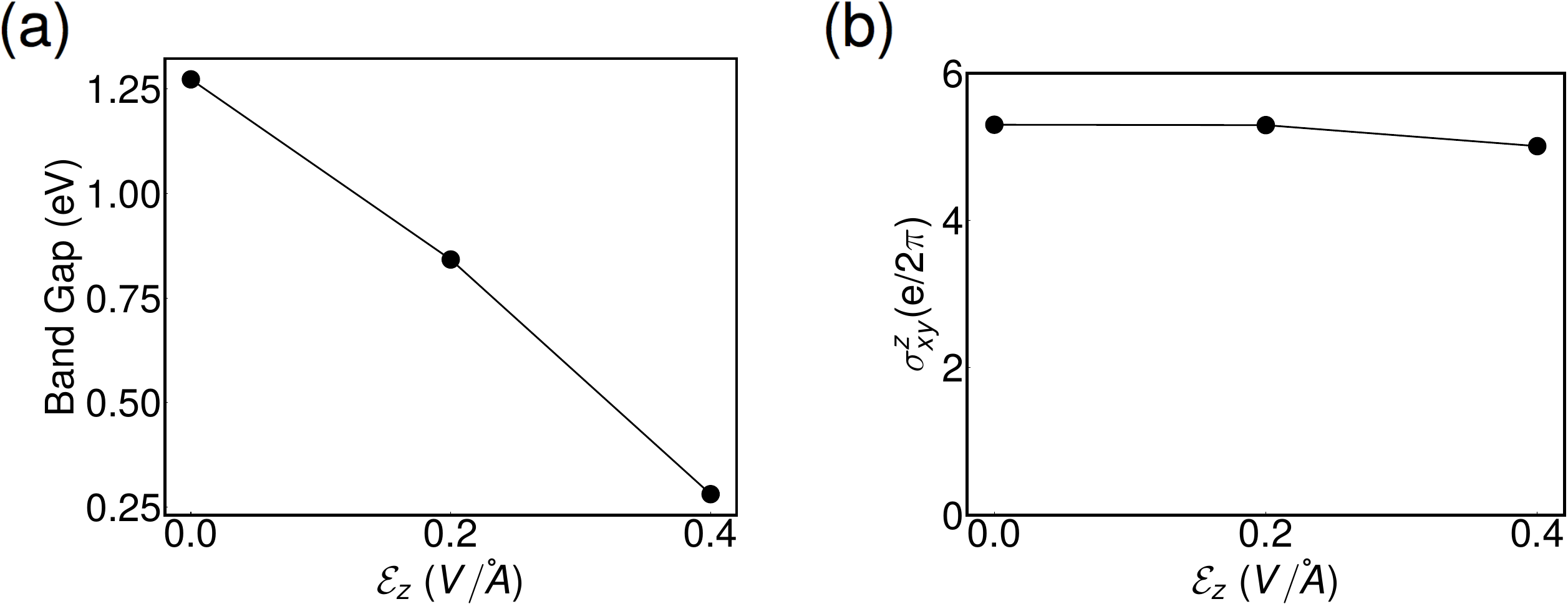}   
  
	\caption{Evolution of the (a) band gap and (b) orbital-Hall conductivity of the MoS$_{2}$ bilayer with an applied perpendicular electrical field $\mathcal{E}_{z}$. }
	\label{fig:fig3s}
\end{figure} 

\section{Angular Momentum Weighted Berry Curvature -  PAOFLOW}
To show the validity of the low-energy model to calculate the OHC in bilayer TMDs,  we can inspect the contribution of the different high symmetry points to the integrand ${\cal Q}(\vec{k})$ of $\sigma^{z}_{OH}$:
\begin{eqnarray}
\sigma^{z}_{OH}\propto  \int_{BZ}  {\cal Q}(\vec{k}) d^2k, ~~{\cal Q}(\vec{k})=\sum_{n}\Theta(E_n-E_F)~\Omega_{n,\vec{k}}^{L_z}.
\label{conductivitys}
\end{eqnarray}

In Fig.~\ref{fig:berry} we show ${\cal Q}(\vec{k})$ along the high symmetry lines of the Brillouin zone.  It has a sharp strong peak in $K$,  showing that although the influence of other symmetry points is relevant for the OHE in this system, the dominant contribution to the OHE plateau arrises from the Berry curvature of the valence band at the K points.  ${\cal Q}(\vec{k})$ also presents  a secondary broad peak in $M$. For monolayer TMDs the contribution from $M$ is reduced~\cite{OHE_Bhowal_1}.

\begin{figure}[h!]
	\centering
	 \includegraphics[width=0.5\linewidth,clip]{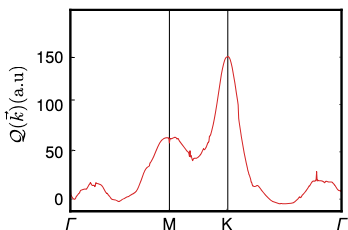}   
  
	\caption{${\cal Q}(\vec{k})$ along the high symmetry lines of the hexagonal Brillouin zone of the bilayer MoS$_{2}$. }
	\label{fig:berry}
\end{figure} 

\section{Orbital Chern number} 
In the main text, we present an orbital Chern number as the topological invariant which indexes the orbital Hall insulating phase of the TMD bilayers ($\mathcal{C}_L=2$). To compute this orbital Chern number, we followed the method introduced by D. N. Sheng, {\it et. al.} for finite systems \cite{Sheng-ChernNumber_2006}, and later formalized by E. Prodan \cite{Prodan-ChernNumber-2009} in the thermodynamic limit. This method has been applied successfully in different situations \cite{Sheng-ChernNumber_2011, Sheng-Isospin-Chern-number} and it is ideal for systems where the operator associated with the Chern number does not commute with the Hamiltonian, as in the case of the orbital angular momentum ($[\tilde{H}(\vec{q}_\tau), L_z]\neq 0$) . 

\begin{enumerate}
\item First, we calculate the energy spectrum and eigenvectors of the Hamiltonian that describes the electronic structure of the system, such as the one given by Eq. (3) of the main text. Since we are interested in the topological properties of insulators, the band structure must be fully gapped. We may then separate the eigenstates of the Hamiltonian that belong to the conduction- and valence-band subspaces. With our effective model for the bilayer, the valence-band subspace is spanned by two eigenstates $\big|\psi_{n,\vec{q}}^{s,\tau} \big> $ that clearly depend upon $s$ and $\tau$. Here, $n =1,2$  label the two valence energy bands, $\vec{q}$ represents the wavevector relative to valleys, $\tau =\pm 1$ denote the valley quantum number associated with the $K$ and $K'$ symmetry points, and $s=\uparrow, \downarrow$ designate the two spin sectors. 


\item The second step consists in projecting the orbital angular momentum operator in the valence band subspace. To this end, we construct the matrix $\left(\mathbb{L}^\text{v}(\vec{k})=P(\vec{k}){L_z}P(\vec{k})\right)$, where $P(\vec{k})$ is the projection operator on the valence band states. The matrix elements of $\mathbb{L}^\text{v}(\vec{k})$ are given by  $\big<\psi_{n,\vec{q}}^{s,\tau}\big|L_z\big|\psi_{m,\vec{q}}^{s,\tau}\big>$ ($n,m=1,2$), which only takes into account the valence band states. We then compute its eigenvalues and eigenvectors. The eigenvalue spectrum of $\mathbb{L}^\text{v}(\vec{q})$ is presented in  Fig. 3 (a) of the main text and it may be separated into positive $\epsilon^+_{s,\tau}(\vec{q})$ and negative $\epsilon^-_{s,\tau}(\vec{q})$ eigenvalues. The application Prodan's method requires that the eigenvalue spectrum must be gapped. 

\item The third step is the construction of the eigenvectors of $\mathbb{L}^\text{v}(\vec{k})$ with the use of the coefficients $\big[\alpha^{\pm}_{s,\tau}(\vec{q}), \beta^{\pm}_{s,\tau}(\vec{q})\big]$ obtained with the diagonalization of $\mathbb{L}^\text{v}(\vec{k})$:
\begin{eqnarray}
\big|\Phi^{\pm}_{s,\tau}\big>=\alpha^{\pm}_{s,\tau}(\vec{q}) \big|\psi_{1,\vec{q}}^{s,\tau}\big> + \beta^{\pm}_{s,\tau}(\vec{q}) \big|\psi_{2,\vec{q}}^{s,\tau}\big>.
\end{eqnarray}

\item The last step is the computation of the orbital Chern number using Eq. (4) of the main text 
\begin{eqnarray}
\mathcal{C}_L^{\pm}=\frac{1}{2\pi}\int d^2q \sum_{s,\tau} F^{\pm}_{s,\tau}(q),
\label{ClSupmat}
\end{eqnarray}
where $F^{\pm}_{s,\tau}(q)=-2 \text{Im} [\big<\partial_{q_x}\Phi_{s,\tau}^{\pm}(\vec{q})\big|\partial_{q_y}\Phi_{s,\tau}^{\pm}(\vec{q})\big>]$. To evaluate the integrals  in Eq. (\ref{ClSupmat}), we have used a cutoff in the momentum space $\Lambda=15 \angstrom^{-1}$, which provides a numerical deviation smaller than $1\%$ from the quantized value $\mathcal{C}_L=(1/2)(\mathcal{C}^+_L-\mathcal{C}^-_L)=2$. Note that this cut-off value is larger than the limit of validity of effective Hamiltonian given by Eq. (3) of the main text. However, this is not an issue in  the context of  the low energy continuous theory since the function $\sum_{s,\tau}F^{\pm}_{s,\tau}(q)$ is strongly peaked in the valleys and decays relatively fast for increasing values of $q$.
\end{enumerate}
\subsection{Limitations of the analysis}
The method discussed above relies on a fully gapped spectrum of the orbital angular momentum operator, when projected into the valence band subspace. This condition is fulfilled for the low-energy Hamiltonian. However, differently from the case of the spin operator in the quantum spin Hall insulator, the spectrum does not remain opened in the whole Brillouin zone. Fortunately, one can still define a topological number, in a similar way of what is done for valley Chern numbers~\cite{Zhang2013}.  Because the $L_z$ projected Berry curvature is concentrated around the valley points, the integral of eq. \ref{ClSupmat} is well defined and we can obtain the orbital Chern number. Rigorously, it is an approximation since it does not consider the whole Brillouin zone. In this sense, our analysis has the same limitations of the quantum valley Hall effect.  For instance, the existence of edge-states is limited to to edges that preserve the two valleys, as in the case of zigzag nanoribbons.  On the other hand, to calculate the OHE, differently from the VHE, it is possible to use the orbitally projected Berry curvature in the whole BZ, which indicates that it might be possible to find a different strategy to calculate the orbital Chern number using the whole BZ.


\end{widetext}

\end{document}